# Numerical methods for spin-dependent transport calculations and spin bound states analysis in Rashba waveguides

Hang Xie[1,2], Feng Jiang[3], Wei E.I. Sha[4]

[1]Department of Applied Physics, Chongqing University, Chongqing, China
[2]Department of Chemistry, the University of Hong Kong, HKSAR, China
[3]School of Mathematics and Physics, Shanghai University of Electric Power, Shanghai, China
[4]Department of Electrical and Electronic Engineering, the University of Hong Kong, HKSAR, China

Email: xiehang2001@hotmail.com

**Abstract**
Numerical methods are developed in the quantum transport calculations for electron in the waveguides with spin-orbital (Rashba) interaction. The methods are based on a hybrid mode-matching scheme in which the wavefunctions are expressed as the superposition of eigenmodes in the lead regions and in the device region the wavefunction is expressed on the discrete basis. Two versions are presented for the lead without and with the Rashba interaction. In the latter case the eigenmodes are obtained from a quadratic eigenproblem calculation. These methods are suitable for the systems with variable geometries or arbitrary potential profiles. The computation can be effectively accelerated by the sparse matrix technique. We also investigate the Fano-Rashba bound states in the Rashba waveguides by some nonlinear eigenstate calculation. This calculation is based on a mode-matching method and self-consistent results are obtained in our calculations.

## 1. Introduction

Spin-dependent electron ballistic transport plays very important roles in the spintronics devices. In 1990s Datta and Das proposed a famous spin version of field effect transistor based on the spin-orbital interaction (SOI) or the Rashba effect [1]. In this Datta-Das-type spin transistor, the spin polarized current enters the SOI region and decomposed into different eigenmodes with different group velocities. Then the output spin components may change due to the interference of these eigenmodes, which is called the spin precession effect [2]. Since the spin polarization ratio injected from ferromagnets into semiconductors is very low in experiment due to the conductivity mismatch [3], some alternative ways are proposed which use the spatially inhomogeneous spin polarization in some T-shaped SOI regions or in some gate-modulated Rashba device [4-6].

To understand the transport properties of the 2-dimentional electron gas (2DEG) in the SOI systems, many calculations methods are proposed. Besides the perturbation method [2] which is used only for the weak SOI coupling, and some analytical method [7] which is only suitable for the parabolic potential, there exist two types of universal methods for exactly calculating the electron transport in SOI systems: one is the nonequilibrium Green's function (NEGF) theory [5-6,8] and the other is the scattering matrix theory [9-11]. The NEGF method introduces the self-energy to account for the influence of electrodes. This method is also suitable for the atomic structures since it uses the tight binding model. The scattering matrix method uses the effective mass approximation for solving the Schrödinger equation. In this method the wavefunctions are expanded into eigenmodes superposition and the relation between the expansion coefficient vectors in two adjacent parts is constructed by the transfer matrix. This method is very suitable for mesoscopic calculations as the main eigenmodes may be of very low energies corresponding to large electron wavelength (such as 100 nm). But this method can only deal with the waveguide-shaped device which can be divided into parallel stripes for the transfer matrix calculations.

To overcome this shortcoming, some hybrid-mode-matching (HMM) method, or the quantum-transmitting-boundary method (QTBM) is developed [12-14]. These methods combine the advantages of the scattering matrix and the NEGF theory: in the lead regions the wavefunction is expanded into a superposition of eigenmodes but in the device region the wavefunction is represented on the discrete basis (it is defined at each grid node; and the finite difference or finite element scheme is used to discretize the Schrödinger equation). So these methods are more suitable for the device with arbitrary geometry and potential profile, compared to the scattering matrix method. In the work of L. Serra and D. Sanchez, a simple extension of QTBM is given to the Rashba system [15]. Here we give a detailed and systematic generalization of the HMM method to spin systems. Two versions of the generalized HMM method are proposed, one is for the lead without SOI, and the other is for the lead with SOI. In the latter case, to obtain the eigenmodes in a uniform SOI lead, the mode expansion and the quadratic eigenstate method are implemented. In these methods not only the propagation modes, but also the evanescent modes are considered. All these modes consist of the two spin components and the transverse motion and longitudinal (in the lead direction) motion are coupled with each other.

In this paper, we also use our methods to investigate some Fano resonance in the SOI quantum well systems. Fano resonance is an important phenomenon that exhibits the interaction between the continuous propagation modes and the discrete bound states in a transport system [16-17]. In the system with SOI, there also exists such resonance called the Fano-Rashba resonance [18-19]. We use our methods to study this phenomenon and find that in some SOI quantum wells there are multiple Fano-Rashba resonances that come from the high-order bound states of the spinor. With a simplified mode matching method we solve out these spin bound states by a nonlinear eigenstate procedure. The energies of these bound states are also related to the Fano-Rashba resonance peaks in the transmission spectra. To our knowledge it is the first time in this paper to give the detailed investigation for these high-order spin bound states in the Fano-Rashba resonance.

The paper is organized as follows. In Section 2 we give the basic theory for the spin-orbital interaction. Section 3 gives the details of the hybrid mode-matching method, including the eigenmodes calculation in the SOI waveguide. Section 4 presents the bound state investigation by a simplified mode-matching method. In Section 5 we show our numerical results and discussions. The conclusion is given in Section 6.

## 2. Basic theory for the Rashba interaction

The device that we study is the 2D-electron gas (2DEG) system, which exists in some semiconductor hetero-structures. The electron wavefunction is restricted in the interface of two types of semiconductors under some triangular potential due to the bend of two energy bands. This restriction potential also provides some electric field perpendicular to the interface ($z$ direction), which gives rise to the Rashba interactions in the 2DEG system: $H_R = \frac{\hbar e}{4m^2c^2}(\vec{\sigma} \times \vec{p}) \cdot \vec{E}$, where $\vec{\sigma}$ is the Pauli matrix, $\vec{p} = -i\hbar\vec{\nabla}$ is the momentum operator, $\vec{E}$ is the electric field applied on the electron gas, $m$ is the electron mass, $e$ is the element charge, $c$ is the light velocity and $\hbar$ is the reduced Planck constant., The Rashba term is also written in a symmetrized form to ensure it to be Hermitian when the SOI factor $\vec{\alpha}(\vec{r})$ is spatially dependent [9,20]

$$H_R = \frac{1}{2\hbar}[\vec{\alpha}(\vec{r}) \cdot (\vec{\sigma} \times \vec{p}) + (\vec{\sigma} \times \vec{p}) \cdot \vec{\alpha}(\vec{r})] \tag{1}$$

where $\vec{\alpha} = \dfrac{\hbar^2 e \vec{E}}{4m^2 c^2}$ is called the Rashba factor. When $\vec{E}$ is in the z direction, and under the effective mass approximation, the Schrödinger equation with the spin-orbital interaction is given below:

$$\{-\dfrac{\hbar^2}{2m}\nabla^2 + V(\vec{r}) + \dfrac{i}{2}[\alpha_z(\vec{r})(\boldsymbol{\sigma}_y \partial_x - \boldsymbol{\sigma}_x \partial_y) + (\boldsymbol{\sigma}_y \partial_x - \boldsymbol{\sigma}_x \partial_y)\alpha_z(\vec{r})]\}\boldsymbol{\psi}(\vec{r}) = E\boldsymbol{\psi}(\vec{r}) \qquad (2)$$

where $V(\vec{r})$ is the potential for the 2DEG, $\boldsymbol{\psi}(\vec{r}) = \begin{pmatrix} \psi^{(1)}(\vec{r}) \\ \psi^{(2)}(\vec{r}) \end{pmatrix}$ is a two-component vector for the wavefunction with spin up and down. In our calculation, the Schrödinger equation above is rewritten as

$$\begin{pmatrix} \nabla^2 + k_0^2 \varepsilon(\vec{r}) & -\overline{\alpha}(\vec{r}) \cdot (\partial_x - i\partial_y) - (\partial_x - i\partial_y) \cdot \overline{\alpha}(\vec{r}) \\ \overline{\alpha}(\vec{r}) \cdot (\partial_x + i\partial_y) + (\partial_x + i\partial_y) \cdot \overline{\alpha}(\vec{r}) & \nabla^2 + k_0^2 \varepsilon(\vec{r}) \end{pmatrix} \begin{pmatrix} \psi^{(1)}(\vec{r}) \\ \psi^{(2)}(\vec{r}) \end{pmatrix} = \begin{pmatrix} 0 \\ 0 \end{pmatrix} \qquad (3)$$

where $k_0^2 \cdot \varepsilon(r) = \dfrac{2mE}{\hbar^2} \cdot [1.0 - \dfrac{V(\vec{r})}{E}]$, $k_0$ is the wavevector of the free electron, $k_0^2 = \dfrac{2mE}{\hbar^2}$, $\varepsilon(r)$ is the equivalent dielectric constant: $\varepsilon(r) = 1.0 - V(\vec{r})/E$; $\overline{\alpha}(\vec{r})$ is the reduced Rashba factor: $\overline{\alpha}(\vec{r}) = \dfrac{m}{\hbar^2}\alpha_z(\vec{r})$.

In the following section, we will introduce the hybrid mode-matching method for the transport calculation.

## 3. Hybrid mode-matching method

In this paper, we extend the hybrid mode-matching method into the spin version. The system is demonstrated in Fig. 1. The device are defined in the areas of $0 \leq x \leq H$ and $0 \leq y \leq a$. The electron leads are defined in the areas of $x < 0$ and $x > H$ with $0 \leq y \leq a$. In the device region, the wavefunctions are expressed on the discrete nodes (on a $N_a$ by $N_b$ mesh).

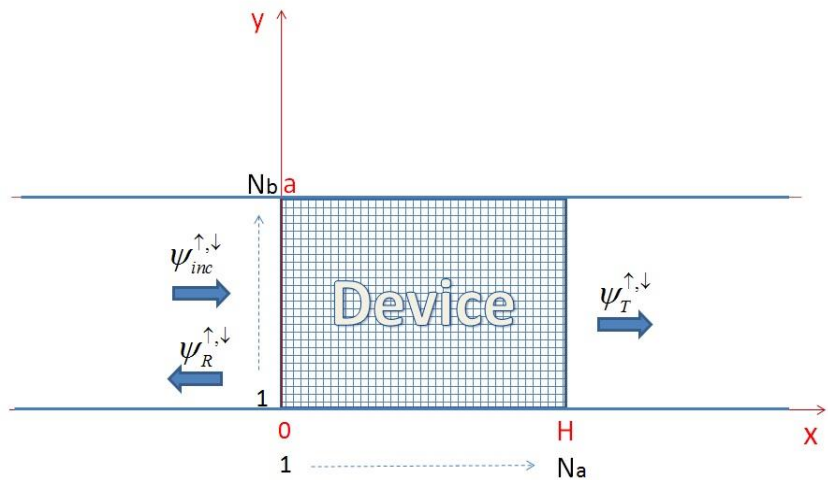

**Fig. 1.** The diagram that demonstrates the hybrid mode-matching method. The left and the right parts are the two electrodes regions; the middle part with grid is the device region. The incident electron wave ($\psi_{inc}^{\uparrow,\downarrow}$) and reflected electron wave ($\psi_R^{\uparrow,\downarrow}$) are indicated in the left lead; the transmitted electron wave ($\psi_T^{\uparrow,\downarrow}$) is indicated in the right lead. The superscripts denote the spin-up and spin-down components of electron.

### 3.1 zero SOI lead

In this case, we assume that only in the middle device region there exists uniform SOI and in the two lead regions the SOI can be neglected by the modulation of the gate voltage. A spin-polarized current is injected into the SOI waveguide region from one ferromagnetic material and perfect ohmic contact between the ferromagnetic material and the semiconductor is assumed.

The Schrödinger equation in the device region is discretized in a middle finite-difference format

$$\nabla^2 \psi^{(1)}(\vec{r}) \approx \frac{\psi^{(1)}_{i+1,j} + \psi^{(1)}_{i-1,j} + \psi^{(1)}_{i,j+1} + \psi^{(1)}_{i,j-1} - 4\psi^{(1)}_{i,j}}{d^2} \quad ,$$

$$\partial_x \psi^{(1)}(\vec{r}) \approx \frac{\psi^{(1)}_{i+1,j} - \psi^{(1)}_{i-1,j}}{2d}, \quad \partial_y \psi^{(1)}(\vec{r}) \approx \frac{\psi^{(1)}_{i,j+1} - \psi^{(1)}_{i,j-1}}{2d}.$$

where $\psi^{(1)}_{i,j}$ ($\psi^{(2)}_{i,j}$) is the discretized wavefunction for the spin-up (down) component with the coordinate $(i,j)$; $d = a/(N_a - 1) = b/(N_b - 1)$, is the size of one unit cell in the device grid. With this format, Eq. (3) can be discretized into the following form

$$\begin{aligned}
&\psi^{(1)}_{i+1,j} + \psi^{(1)}_{i-1,j} + \psi^{(1)}_{i,j+1} + \psi^{(1)}_{i,j-1} - 4\psi^{(1)}_{i,j} + k_0^2 d^2 \varepsilon_{i,j} \cdot \psi^{(1)}_{i,j} \\
&- \overline{\alpha}_{i,j} \frac{d}{2} [\psi^{(2)}_{i+1,j} - \psi^{(2)}_{i-1,j}] - \frac{d}{2} [\overline{\alpha}_{i+1,j} \psi^{(2)}_{i+1,j} - \overline{\alpha}_{i-1,j} \psi^{(2)}_{i-1,j}] \\
&+ i\overline{\alpha}_{i,j} \frac{d}{2} [\psi^{(2)}_{i,j+1} - \psi^{(2)}_{i,j-1}] + i \frac{d}{2} [\overline{\alpha}_{i,j+1} \psi^{(2)}_{i,j+1} - \overline{\alpha}_{i,j-1} \psi^{(2)}_{i,j-1}] = 0
\end{aligned} \quad (4a)$$

$$\begin{aligned}
&\psi^{(2)}_{i+1,j} + \psi^{(2)}_{i-1,j} + \psi^{(2)}_{i,j+1} + \psi^{(2)}_{i,j-1} - 4\psi^{(2)}_{i,j} + k_0^2 d^2 \varepsilon_{i,j} \cdot \psi^{(2)}_{i,j} \\
&+ \overline{\alpha}_{i,j} \frac{d}{2} [\psi^{(1)}_{i+1,j} - \psi^{(1)}_{i-1,j}] + \frac{d}{2} [\overline{\alpha}_{i+1,j} \psi^{(1)}_{i+1,j} - \overline{\alpha}_{i-1,j} \psi^{(1)}_{i-1,j}] \\
&+ i\overline{\alpha}_{i,j} \frac{d}{2} [\psi^{(1)}_{i,j+1} - \psi^{(1)}_{i,j-1}] + i \frac{d}{2} [\overline{\alpha}_{i,j+1} \psi^{(1)}_{i,j+1} - \overline{\alpha}_{i,j-1} \psi^{(1)}_{i,j-1}] = 0
\end{aligned} \quad (4b)$$

Similar to the reference [12], we assume there is a hard-wall potential in the $y$ direction. Then the wavefunction at the boundary of $y = 0$ and $y = a$ is zero. This makes the discretized wavefunctions at $j=1$ and $j=N_b$ to be zero. So there are totally $2N_a(N_b - 2)$ unknowns for the free $\psi^{(1)}_{i,j}$ and $\psi^{(2)}_{i,j}$; and there are $2(N_a - 2)(N_b - 2)$ independent equations for Eq. (4a) and Eq. (4b).

For the wavefunctions in the lead regions, we decompose them into a series of eigenmodes. Here we assume that the lead regions are uniform in the propagation direction and have no SO interaction. The eigenmode has the form of $e^{\pm ik_n x} \phi_n(y)$, where $\phi_n(y)$ is the eigenfunction of the following equation

$$[-\frac{\hbar^2}{2m} \frac{\partial^2}{\partial y^2} + V(y)] \phi(y) = \varepsilon_n \phi(y).$$

For simplicity, we also assume that the lead has a hard-wall potential in the transverse direction and it is easy to see that $\phi_n(y) = \sin(\frac{n\pi}{a} y)$, so $\varepsilon_n = \frac{\hbar^2}{2m}(\frac{n\pi}{a})^2$ and $k_n = \sqrt{\frac{2m}{\hbar^2} E - (\frac{n\pi}{a})^2}$. In the left lead region we assume there is an incident wave with the eigen-mode index ($n_0$) and amplitudes ($A_{inc}^\uparrow$, $A_{inc}^\downarrow$). The wavefunctions in the left and right leads are thus expressed as

$$\psi_{left}^{\uparrow,\downarrow}(x,y) = A_{inc}^{\uparrow,\downarrow} e^{ik_{n_0} x} \sin(\frac{n_0 \pi}{a} y) + \sum_{n=1}^{N_1} A_n^{\uparrow,\downarrow} e^{-ik_n x} \sin(\frac{n\pi}{a} y) \qquad (5)$$

$$\psi_{right}^{\uparrow,\downarrow}(x,y) = \sum_{n=1}^{N_1} B_n^{\uparrow,\downarrow} e^{ik_n (x-H)} \sin(\frac{n\pi}{a} y), \qquad (6)$$

where $A_n^{\uparrow,\downarrow}$ and $B_n^{\uparrow,\downarrow}$ are the expansion coefficients. From the discrete sine transformation, we set $N_1 = N_b - 2$, where $N_1$ is the number of the free grid points in the device-lead interface. It is noted that when $n$ is large enough, $k_n$ becomes pure imaginary. So here we also include the evanescent lead modes.

Then we implement the continuity condition to relate the wavefunctions in the leads and device regions. This condition means the wavefunction and the current flux are continuous across any boundary. It is easy to see that if the Rashba factor ($\vec{\alpha}(\vec{r})$) varies smoothly, the flux continuity is identical to the derivative continuity of the wavefunctions. (If the Rashba factor abruptly changes in some piecewise system, there exist the derivative discontinuity [20] and additional term should be accounted. see Sec. 4 for details). Here we assume a smoothly change of Rashba factor from zero in the leads to a finite value in the device [21]. So the continuity conditions for both the wavefunctions and their normal derivatives can be implemented on the boundaries. The detailed results are shown in the following.

(1) At the boundary between the left-lead and the device, we have

$$\psi_{left}^{\uparrow,\downarrow}|_{x=0} = \psi_{1,j}^{(1),(2)}, \qquad (7)$$

$$(\partial_x \psi_{left}^{\uparrow,\downarrow})|_{x=\frac{d}{2}} = \frac{\psi_{2,j}^{(1),(2)} - \psi_{1,j}^{(1),(2)}}{d}. \qquad (8)$$

We set the derivative continuity condition at $x=d/2$ because the finite difference $[\psi_{2,j}^{(1),(2)} - \psi_{1,j}^{(1),(2)}]/d$ is the approximant of $\partial_x \psi_{left}^{\uparrow,\downarrow}(x,y)$ at $x=0$ with the first-order accuracy, but it also the approximant of $\partial_x \psi_{left}^{\uparrow,\downarrow}(x,y)$ at $x=d/2$ with the second-order accuracy [12]. So choosing the derivative continuity condition at $x=d/2$ can lead better precision. Substituting Eq. (5) into Eqs. (7)-(8), we get

$$A_{inc}^\uparrow \sin(\frac{n_0 \pi}{a} y_j) + \sum_{n=1}^{N_1} A_n^\uparrow \sin(\frac{n\pi}{a} y_j) = \psi_{1,j}^{(1)}, \qquad (9a)$$

$$A_{inc}^\downarrow \sin(\frac{n_0 \pi}{a} y_j) + \sum_{n=1}^{N_1} A_n^\downarrow \sin(\frac{n\pi}{a} y_j) = \psi_{1,j}^{(2)}, \qquad (9b)$$

$$A^{\uparrow}_{inc}(ik_{n_0})e^{ik_{n_0}(d/2)}\sin(\frac{n_0\pi}{a}y_j)+\sum_{n=1}^{N_1}A^{\uparrow}_n e^{-ik_n(d/2)}(ik_n)\sin(\frac{n\pi}{a}y_j)=\frac{\psi^{(1)}_{2,j}-\psi^{(1)}_{1,j}}{d}, \tag{10a}$$

$$A^{\downarrow}_{inc}(ik_{n_0})e^{ik_{n_0}(d/2)}\sin(\frac{n_0\pi}{a}y_j)+\sum_{n=1}^{N_1}A^{\downarrow}_n e^{-ik_n(d/2)}(ik_n)\sin(\frac{n\pi}{a}y_j)=\frac{\psi^{(2)}_{2,j}-\psi^{(2)}_{1,j}}{d}. \tag{10b}$$

Since we only compare the quantities on the free nodes at the boundary, $2\leq j\leq N_b$, we have $4(N_b-2)$ independent equations at this boundary.

(2) Similarly, at the boundary between the device and the right lead, we have the following relations

$$\sum_{n=1}^{N_1}B^{\uparrow}_n\sin(\frac{n\pi}{a}y_j)=\psi^{(1)}_{N_a,j}, \tag{11a}$$

$$\sum_{n=1}^{N_1}B^{\downarrow}_n\sin(\frac{n\pi}{a}y_j)=\psi^{(2)}_{N_a,j}, \tag{11b}$$

$$\sum_{n=1}^{N_1}B^{\uparrow}_n e^{ik_n(-d/2)}(ik_n)\sin(\frac{n\pi}{a}y_j)=\frac{\psi^{(1)}_{N_a,j}-\psi^{(1)}_{N_a-1,j}}{d}, \tag{12a}$$

$$\sum_{n=1}^{N_1}B^{\downarrow}_n e^{ik_n(-d/2)}(ik_n)\sin(\frac{n\pi}{a}y_j)=\frac{\psi^{(2)}_{N_a,j}-\psi^{(2)}_{N_a-1,j}}{d}. \tag{12b}$$

Again we have $4(N_b-2)$ independent equations at this boundary. It is easy to obtain the total number of equations for this HMM method: $N_f=2(N_b-2)(N_a-2)+8(N_b-2)$, and the number of unknowns $N_x=2(N_b-2)N_a+4(N_b-2)$. We can check that $N_x=N_f$.

After assembling all these equations into a large matrix equation, we may solve it numerically. In the case of a small grid, the common linear algebra package like Lapack can be used to solve this matrix equation. But when the grid size is large, the large memory and long computation time lead to very heavy calculations. Since most of the equations are from the finite difference equations of the device, the large matrix is very sparse. So the sparse matrix technique is used to solve these linear equations with very high efficiency [12-13, 22].

When all the discrete points and expansion coefficients are solved out, the transmitted current on the right lead region is calculated as an integral of the current flux on the transverse ($y$) direction

$$I^{\uparrow,\downarrow}_T=\int_0^a J^{\uparrow,\downarrow}_{T,x}(y)dy, \tag{13}$$

where $J^{\uparrow}_{T,x}=\frac{\hbar}{m}\text{Im}[\psi^{(1)*}\partial_x\psi^{(1)}]$; $J^{\downarrow}_{T,x}=\frac{\hbar}{m}\text{Im}[\psi^{(2)*}\partial_x\psi^{(2)}]$. For simplicity, we only consider the up-polarized incident wave, thus $I_{inc}=I^{\uparrow}_{inc}$. Then we obtain the transmission coefficient by

$$T^{\uparrow,\downarrow}=I^{\uparrow,\downarrow}_{T,x}/I_{inc}. \tag{14}$$

3.2. Eigenmodes in the SOI lead

If we assume the lead regions also have the Rashba interaction, we have to use the eigenmodes in these SOI

leads. These eigenmodes are assumed to have the guiding-wave form with two spin components as

$$\psi(x,y) = \begin{pmatrix} \Phi^\uparrow(y) \\ \Phi^\downarrow(y) \end{pmatrix} e^{ik_x \cdot x}, \tag{15}$$

Substituting above into Eq. (2), and noticing that in a uniform SOI waveguide a constant SOI factor reduces the symmetrized Hamiltonian (Eq.(3)) into the following one,

$$\begin{pmatrix} -\frac{\hbar^2}{2m}\nabla^2 + V & \alpha(\partial_x - i\partial_y) \\ \alpha(-\partial_x - i\partial_y) & -\frac{\hbar^2}{2m}\nabla^2 + V \end{pmatrix} \begin{pmatrix} e^{ik_x x}\Phi^\uparrow(y) \\ e^{ik_x x}\Phi^\downarrow(y) \end{pmatrix} = E \begin{pmatrix} e^{ik_x x}\Phi^\uparrow(y) \\ e^{ik_x x}\Phi^\downarrow(y) \end{pmatrix}. \tag{16}$$

If we use some basis set {$\phi_n(y)$} to expand the functions $\Phi_m^{\uparrow,\downarrow}(y)$ as

$$\Phi^\uparrow(y) = \sum_{n=1}^N a_n \phi_n(y); \quad \Phi^\downarrow(y) = \sum_{n=1}^N b_n \phi_n(y),$$

and substituting them into Eq. (12), we have the following two sets of equations

$$\sum_n^N a_n[(-\frac{\hbar^2}{2m}\partial_y^2 + V(y)) + \frac{\hbar^2}{2m}k_x^2]\phi_n(y)e^{ik_x x} + \sum_n^N b_n \alpha(ik_x - i\partial_y)\phi_n(y)e^{ik_x x} = E\sum_n^N a_n \phi_n(y)e^{ik_x x} \tag{17a}$$

$$\sum_n^N a_n \alpha(-ik_x - i\partial_y)\phi_n(y)e^{ik_x x} + \sum_n^N b_n[(-\frac{\hbar^2}{2m}\partial_y^2 + V(y)) + \frac{\hbar^2}{2m}k_x^2]\phi_n(y)e^{ik_x x} = E\sum_n^N b_n \phi_n(y)e^{ik_x x}. \tag{17b}$$

By multiplying $\phi_m^*(y)$ on the two sides of these equations and doing the integral ($\int dy \phi_m^*(y)$), and noticing the following relations

$$[\frac{-\hbar^2}{2m}\partial_y^2 + V(y)]\phi_n(y) = \varepsilon_n \phi_n(y), \tag{18}$$

$$\int_0^a \phi_m^*(y)\phi_n(y)dy = \delta_{mn}, \tag{19}$$

we obtain the equations for $\{a_n\}$ and $\{b_n\}$

$$\sum_n^N a_n[\varepsilon_n + \frac{\hbar^2}{2m}k_x^2]\delta_{mn} + \sum_n^N b_n \alpha(ik_x - i<m|\partial_y|n>) = E\sum_n^N a_n \delta_{mn}, \tag{20a}$$

$$\sum_n^N a_n \alpha(-ik_x - i<m|\partial_y|n>) + \sum_n^N b_n[\varepsilon_n + \frac{\hbar^2}{2m}k_x^2]\delta_{mn} = E\sum_n^N b_n \delta_{mn}, \tag{20b}$$

where $<m|\partial_y|n> = \int_0^a \phi_m^*(y)\partial_y \phi_n(y)dy$. In the case of the hard-well potential, $\phi_n(y) = \sqrt{\frac{2}{a}}\sin(\frac{n\pi}{a}y)$; $\varepsilon_n = \frac{\hbar^2}{2m}(\frac{n\pi}{a}y)^2$, and all these matrix elements may be evaluated analytically.

This is a quadratic eigenvalue problem for the Bloch wavevector $k_x$. To solve this eigenproblem, we have to increase the dimension of this matrix equation to transform it into a linear eigenvalue problem [23]. By introducing the following auxiliary vectors: $\mathbf{a}_1 = k_x \mathbf{a}, \mathbf{b}_1 = k_x \mathbf{b}$, ($\mathbf{a}$ and $\mathbf{b}$ are the vector form of expansion coefficients $\{a_n\}$ and $\{b_n\}$) and with the new reduced quantities: $\frac{2mE}{\hbar^2} = k_0^2 = 2\overline{E}$, $\frac{2m\varepsilon_n}{\hbar^2} = 2\overline{\varepsilon}_n$, $\frac{2m\alpha}{\hbar^2} = 2\overline{\alpha}$, the following matrix equations can be derived

$$\begin{pmatrix} \frac{k_x}{2}\mathbf{I} & -\frac{\mathbf{I}}{2} & 0 & 0 \\ (\overline{\varepsilon}_n - \overline{E})\mathbf{I} & \frac{k_x}{2}\mathbf{I} & -i\overline{\alpha}\mathbf{D} & i\overline{\alpha}\mathbf{I} \\ 0 & 0 & \frac{k_x}{2}\mathbf{I} & -\frac{\mathbf{I}}{2} \\ -i\overline{\alpha}\mathbf{D} & -i\overline{\alpha}\mathbf{I} & (\overline{\varepsilon}_n - \overline{E})\mathbf{I} & \frac{k_x}{2}\mathbf{I} \end{pmatrix} \cdot \begin{pmatrix} \mathbf{a} \\ \mathbf{a}_1 \\ \mathbf{b} \\ \mathbf{b}_1 \end{pmatrix} = \begin{pmatrix} 0 \\ 0 \\ 0 \\ 0 \end{pmatrix} \quad (21)$$

where the matrix element $D_{mn} = <m|\partial_y|n>$. It is easy to see that this is an ordinary linear eigenvalue problem with the eigenvalue $\lambda_m$. So we have $k_{x,m} = -2\lambda_m$. With this method, we successfully solve out all the wavevectors $\{k_{x,m}\}$ and eigenvectors for a given energy $E$. Thus the band structure and the eigenfunctions $\Phi_m^{\uparrow,\downarrow}(y)$ for the spin-orbital coupling electron waveguide are obtained.

In practical transport calculations, we have to classify these eigen-wavevectors and the corresponding eigenstates into two subgroups: one is with positive real part for the pure real numbers (propagation modes in the $x+$ direction) or with positive imaginary part for the complex numbers (evanescent modes which decay in the $x+$ direction). They are denoted as $K_m^+$ and $\Phi_m^{+,\uparrow(\downarrow)}(y)$; another is with negative real part for the pure real numbers (propagation modes in the $x-$ direction) or with negative imaginary part for the complex numbers (evanescent modes which decay in the $x-$ direction). The latter are denoted as $K_m^-$ and $\Phi_m^{-,\uparrow(\downarrow)}(y)$. This classification is based on the fact that in the mode expansion process we have to know the direction of these propagation modes and ensure all the evanescent modes do not diverge in the semi-infinite lead regions.

We also notice that L. Serra et.al used the finite difference method, instead of our basis expansion method, to solve out the eigenmodes in the Rashba lead [24]. Their lead has the parabolic potential profile and the eigenmodes are not restricted in the finite transverse region. So some numerical technique such as the complex root searching for a smooth wavefunction derivative was implemented in those eigenmodes calculations. And in our SOI bound state calculations (see Sec. 4 below) we sweep the energy in a range to check if all the secular equations are satisfied. We find our scheme is also similar with theirs for these nonlinear eigenmodes calculations.

3.3 nonzero SOI lead

In this part we will use the SOI eigenmodes to expand the wavefunctions in the lead regions. From the previous section, we see the eigenmodes have the form as $\begin{pmatrix} \Phi_m^{\pm,\uparrow}(y) \\ \Phi_m^{\pm,\downarrow}(y) \end{pmatrix} e^{iK_m^\pm \cdot x}$. Here the spin-up and spin-down components are not independent as they are combined into an eigenmode vector or the so-called spinor. And different from the modes ($\sin(\frac{n\pi}{a} y_j)$) in Sec. 3.1, there are $2N_1$ eigenmodes for these spinors. So in the left and right lead regions, the wave functions are written as,

$$\phi_{left}(x,y) = \begin{pmatrix} \sum_{m}^{2N_1} A_m \Phi_m^{-,\uparrow}(y) \cdot e^{iK_m^- x} \\ \sum_{m}^{2N_1} A_m \Phi_m^{-,\downarrow}(y) \cdot e^{iK_m^- x} \end{pmatrix} + \begin{pmatrix} A_{inc} \Phi_{m_0}^{+,\uparrow}(y) \cdot e^{iK_{m_0}^+ x} \\ A_{inc} \Phi_{m_0}^{+,\downarrow}(y) \cdot e^{iK_{m_0}^+ x} \end{pmatrix}, \quad (22)$$

$$\phi_{right}(x,y) = \begin{pmatrix} \sum_{m}^{2N_1} B_m \Phi_m^{+,\uparrow}(y) \cdot e^{iK_m^+ (x-H)} \\ \sum_{m}^{2N_1} B_m \Phi_m^{+,\downarrow}(y) \cdot e^{iK_m^+ (x-H)} \end{pmatrix}. \quad (23)$$

Here the number of unknowns are $N_x = 2N_a N_1 + 4N_1$, where for $A_m$ and $B_m$ each of them has the number of $2N_1$.

Then we use the continuity conditions to relate the wavefunctions (and their derivatives) in the device and lead regions. Similar equations can be written as those in Eqs.(9)-(12), with $\Phi_m^{\pm,\uparrow,\downarrow}(y_j)$ and $K_m^\pm$ to replace $\sin(\frac{n\pi}{a} y_j)$ and $\pm k_m$. (Here we see in the case of Rashba effect, the degeneracy of transverse eigenmode ($\sin(\frac{n\pi}{a} y_j)$) is broken.) From these formulas, we see there are $8N_1$ connection equations on the boundaries. Including the equations in the device region, there are totally $N_f = 2N_1(N_a - 2) + 8N_1$ equations. So we also have the relation $N_x = N_f$. For the large system, the sparse matrix technique can also be implemented for solving these matrix equations.

After solving out all the unknowns, we can calculate the current in this SOI system. But for this SOI system, the current flux is not spin-independent. It includes all the spin components. The spinor flux expression may be derived from the Schrödinger equation based on the electron-charge conservation [20]

$$\vec{J} = \frac{\hbar}{m} \text{Re}[-i\psi^\dagger \vec{\nabla}\psi + \frac{m\alpha}{\hbar^2} \psi^\dagger (\hat{z} \times \vec{\sigma})\psi], \quad (24)$$

where $\psi = \begin{Bmatrix} \psi^\uparrow \\ \psi^\downarrow \end{Bmatrix}$ is the spinor and the Rashba vector ($\vec{\alpha} = \alpha\hat{z}$) is in the z direction. We can also written this into the spin-resolved form as

$$J_x = \frac{\hbar}{m}[\text{Im}(\psi^{\uparrow*}\partial_x \psi^\uparrow + \psi^{\downarrow*}\partial_x \psi^\downarrow) + \text{Im}(2\bar{\alpha}\psi^\uparrow \psi^{\downarrow*})], \quad (25a)$$

$$J_y = \frac{\hbar}{m}[\text{Im}(\psi^{\uparrow*}\partial_y \psi^\uparrow + \psi^{\downarrow*}\partial_y \psi^\downarrow) + \text{Re}(2\bar{\alpha}\psi^\uparrow \psi^{\downarrow*})]. \quad (25b)$$

## 4. Spin bound states calculations

In this paper, we will investigate the Fano resonance in SOI waveguides which results from the interaction of continuous propagation modes and the discrete bound states. To see the details of such spin bound state, we develop a calculation scheme based on the traditional mode-matching method. Here we demonstrate the details.

Since these states are bounded in the SOI region, on the non-SOI regions they behave as the evanescent waves. The system geometry is similar with that in Fig. 1. There are three uniform regions in the system: the two lead regions and the device region. Only in the device region there exists a constant Rashba interaction. The energy of the bound state is assumed to lie in the range $E_1$ and $E_2$ ($E_1$ and $E_2$ are the cut-off energy for the 1st and 2nd subband in the non-SOI waveguide). Then we decompose the bound state into $M$ subbands (or eigenmodes). The wavefunctions in different regions are written as follows.

(1) In the left region,

$$\phi_{left}(x,y) = \sum_{n=2}^{M} A_n^{\uparrow} \begin{pmatrix} \sin(\frac{n\pi y}{a}) \\ 0 \end{pmatrix} e^{-ik_n x} + \sum_{n=2}^{M} A_n^{\downarrow} \begin{pmatrix} 0 \\ \sin(\frac{n\pi y}{a}) \end{pmatrix} e^{-ik_n x} . \quad (26)$$

where $k_n$ is given in Sec. 3.1. It is noted that in the left and right regions, $n \geq 2$. This is because that for the bound states with the energy $E \in [E_1, E_2]$, all the expansion modes in these regions are the evanescent modes. So $k_n$ in above equations are pure imaginary numbers.

(2) In the middle SOI device region,

$$\phi_{middle}(x,y) = \sum_{n=1}^{M'} C_n \begin{pmatrix} \Phi_n^{+,\uparrow}(y) \\ \Phi_n^{+,\downarrow}(y) \end{pmatrix} e^{iK_n^+ x} + \sum_{n=1}^{M'} D_n \begin{pmatrix} \Phi_n^{-,\uparrow}(y) \\ \Phi_n^{-,\downarrow}(y) \end{pmatrix} e^{iK_n^- x} . \quad (27)$$

where $\Phi_n^{\pm,\uparrow,\downarrow}(y)$ and $K_n^{\pm}$ come from the band structure calculations for a uniform SOI waveguide provided that the energy $E$ is given (see Sec. 3.2). Here both the propagation modes and the evanescent modes are included. The total number of the eigenmodes is $M'$, which will be determined later.

(3) In the right region,

$$\phi_{right}(x,y) = \sum_{n=2}^{M} B_n^{\uparrow} \begin{pmatrix} \sin(\frac{n\pi y}{a}) \\ 0 \end{pmatrix} e^{ik_n x} + \sum_{n=2}^{M} B_n^{\downarrow} \begin{pmatrix} 0 \\ \sin(\frac{n\pi y}{a}) \end{pmatrix} e^{ik_n x} . \quad (28)$$

To solve this eigenproblem, we set up the secular equations, which come from the continuity conditions. For example, at the boundary between the left lead and the device, the continuity of the wavefunction leads the following equation

$$\sum_{n=1}^{M'} C_n \begin{pmatrix} \Phi_n^{+,\uparrow}(y) \\ \Phi_n^{+,\downarrow}(y) \end{pmatrix} \cdot 1 + \sum_{n=1}^{M'} D_n \begin{pmatrix} \Phi_n^{-,\uparrow}(y) \\ \Phi_n^{-,\downarrow}(y) \end{pmatrix} \cdot 1 = \sum_{n=2}^{M} A_n^{\uparrow} \begin{pmatrix} \sin(\frac{n\pi y}{a}) \\ 0 \end{pmatrix} \cdot 1 + \sum_{n=2}^{M} A_n^{\downarrow} \begin{pmatrix} 0 \\ \sin(\frac{n\pi y}{a}) \end{pmatrix} \cdot 1$$

We multiply $\frac{1}{a}\sin(\frac{n'\pi y}{a})$ on the two sides of the equation above ($2 \leq n' \leq M$), and do the integral $\frac{1}{a}\int_0^a dy$ with the orthogonal relation of $\{\sin(\frac{n\pi y}{a})\}$

$$\frac{1}{a}\int_0^a \sin(\frac{m\pi y}{a})\sin(\frac{n\pi y}{a})dy = \frac{1}{2}\delta_{m,n}, \qquad (29)$$

the following equations are obtained,

$$\sum_{n=1}^{M'} C_n H_{n'n}^{+,\uparrow} + \sum_{n=1}^{M'} D_n H_{n'n}^{-,\uparrow} = \sum_{n=2}^{M} A_n^{\uparrow} \frac{1}{2}\delta_{n',n}, \qquad (30a)$$

$$\sum_{n=1}^{M'} C_n H_{n'n}^{+,\downarrow} + \sum_{n=1}^{M'} D_n H_{n'n}^{-,\downarrow} = \sum_{n=2}^{M} A_n^{\downarrow} \frac{1}{2}\delta_{n',n}, \qquad (30b)$$

where $H_{n'n}^{+(-),\uparrow(\downarrow)}$ are the integrals, defined as follows,

$$H_{n'n}^{+,\uparrow} = \frac{1}{a}\int_0^a \sin(\frac{n'\pi}{a}y)\cdot\Phi_n^{+,\uparrow}(y)dy;\quad H_{n'n}^{-,\uparrow} = \frac{1}{a}\int_0^a \sin(\frac{n'\pi}{a}y)\cdot\Phi_n^{-,\uparrow}(y)dy;$$

$$H_{n'n}^{+,\downarrow} = \frac{1}{a}\int_0^a \sin(\frac{n'\pi}{a}y)\cdot\Phi_n^{+,\downarrow}(y)dy;\quad H_{n'n}^{-,\downarrow} = \frac{1}{a}\int_0^a \sin(\frac{n'\pi}{a}y)\cdot\Phi_n^{-,\downarrow}(y)dy.$$

However, the normal wavefunction derivatives in this piecewise system are not continuous. Only the spinor flux across the system is continuous. If the Rashba factor $\alpha$ is *x*-dependent and changes abruptly at *x*=0 ($\bar{\alpha}(x) = \bar{\alpha}_0 \Theta(x)$, where $\Theta(x)$ is the Heaviside step function.), we have $J_x(0^+) = J_x(0^-)$. From the spinor flux expression (Eq. (24)), the following relation can be derived [20]

$$i\partial_x \psi(0^+) + \bar{\alpha}(0^+)\sigma_y \psi(0^+) = i\partial_x \psi(0^-) + \bar{\alpha}(0^-)\sigma_y \psi(0^-)$$

with the discontinuity of $\alpha$, we have

$$\begin{pmatrix} \partial_x \psi^{\uparrow}(0^+) \\ \partial_x \psi^{\downarrow}(0^+) \end{pmatrix} - \begin{pmatrix} \partial_x \psi^{\uparrow}(0^-) \\ \partial_x \psi^{\downarrow}(0^-) \end{pmatrix} = \bar{\alpha}_0(0^+)\begin{pmatrix} \psi^{\uparrow}(0) \\ -\psi^{\downarrow}(0) \end{pmatrix}.$$

Substituting Eqs. (26) and (27) into the formula above, the following equations hold

$$\sum_{n=1}^{M'} C_n H_{n'n}^{+,\uparrow}\cdot(iK_n^+) + \sum_{n=1}^{M'} D_n H_{n'n}^{-,\uparrow}\cdot(iK_n^-) - \sum_{n=2}^{M} A_n^{\uparrow}\frac{1}{2}\delta_{n',n}\cdot(-ik_n) = \bar{\alpha}_0 \sum_{n=2}^{M} A_n^{\downarrow}\frac{1}{2}\delta_{n',n} \qquad (31a)$$

$$\sum_{n=1}^{M'} C_n H_{n'n}^{+,\downarrow}\cdot(iK_n^+) + \sum_{n=1}^{M'} D_n H_{n'n}^{-,\downarrow}\cdot(iK_n^-) - \sum_{n=2}^{M} A_n^{\downarrow}\frac{1}{2}\delta_{n',n}\cdot(-ik_n) = -\bar{\alpha}_0 \sum_{n=2}^{M} A_n^{\uparrow}\frac{1}{2}\delta_{n',n}. \qquad (31b)$$

We notice that in the work of L. Serra et.al., the eigenmodes are applied similarly as the expansion basis for a potential-step system with SOI as well [24]. But as they calculated for the transmission with evanescent modes, only the linear equations need to be solved, instead of the eigenstates calculation we discuss here.

By the same way, we set up the continuity equations at the boundary between the device and the right lead,

$$\sum_{n=1}^{M'} C_n H_{n'n}^{+,\uparrow} + \sum_{n=1}^{M'} D_n H_{n'n}^{-,\uparrow} = \sum_{n=2}^{M} B_n^{\uparrow}\frac{1}{2}\delta_{n',n} \qquad (32a)$$

$$\sum_{n=1}^{M'} C_n H_{n'n}^{+,\downarrow} + \sum_{n=1}^{M'} D_n H_{n'n}^{-,\downarrow} = \sum_{n=2}^{M} B_n^{\downarrow}\frac{1}{2}\delta_{n',n} \qquad (32b)$$

$$\sum_{n=1}^{M'} C_n H_{n'n}^{+,\uparrow}\cdot(iK_n^+) + \sum_{n=1}^{M'} D_n H_{n'n}^{-,\uparrow}\cdot(iK_n^-) - \sum_{n=2}^{M} B_n^{\uparrow}\frac{1}{2}\delta_{n',n}\cdot(ik_n) = \bar{\alpha}_0 \sum_{n=2}^{M} B_n^{\downarrow}\frac{1}{2}\delta_{n',n} \qquad (33a)$$

$$\sum_{n=1}^{M'} C_n H_{n'n}^{+,\downarrow} \cdot (iK_n^+) + \sum_{n=1}^{M'} D_n H_{n'n}^{-,\downarrow} \cdot (iK_n^-) - \sum_{n=2}^{M} B_n^{\downarrow} \frac{1}{2} \delta_{n',n} \cdot (ik_n) = -\bar{\alpha}_0 \sum_{n=2}^{M} B_n^{\uparrow} \frac{1}{2} \delta_{n',n} \,. \tag{33b}$$

We find that when $M' = 2M - 2$, the number of equations ($N_f$) is equal to the number of expansion coefficients $N_x = 8M - 8$ (see Reference [25]). Thus all the equations above constitute the secular equations for the eigenvalue problem. It is noted that the eigen-energy $E$ is non-linearly dependent on $k_n$, $K_n^{\pm}$ and $\Phi_n^{\pm,\uparrow,\downarrow}(y)$ (or $H_{n'n}^{+(-),\uparrow(\downarrow)}$). So this is an implicit non-linear eigenvalue problem which has to be solved numerically. Here we give one process of such calculation: At first we scan every energy point in the range $[E_1, E_2]$. Since the determinant in Eqs. (30-33) has to be zero at the eigenvalues, we set one coefficient (such as $A_2^{\uparrow}$) to 1.0, and calculated all the other expansion coefficients in Eqs. (30-33) except for one equation (such as the last equation in Eqs. (33b)). The new number of equations (and unknowns) at this time is $N_f - 1 = 8M - 9$. After solving out these equations (or the expansion coefficients), we calculate the error for that unsolved equation with the calculated expansion coefficients (see $\Delta$ in Fig. 6(a) below). The energy that minimizes this error is regarded as the eigenvalue for this bound-state problem.

## 5. Numerical results and discussions

The system in our investigation is the 2DEG defined in a semiconductor heterostructure. The electrons are confined by an approximately triangular potential (in the z direction) perpendicular to the interface plane and can move on the plane. With some other transverse confining potential set up on the 2DEG plane, a 2D electron waveguide forms. The triangular potential has a non-zero gradient of the electric field in the z direction, thus it gives rise to the Rashba spin-orbital interaction. This SOI coupling strength can further be modulated by the gate voltage with some gate mounted on the top of the heterostructure [26-27]. This gate voltage modulation often generates a small potential shift in the SOI region. But in the Sections 5.1-5.4 we negative this potential since it plays a small role in the spin precession process.

5.1 HMM calculations for the lead with zero SOI

The HMM method described in Sec. 3.1 is implemented here. The width ($a$) and length ($H$) of the waveguide are set to 100 nm and 155 nm. In the heterostructure, the effective mass model is used by $m = 0.042 m_e$, where $m_e$ is the free electron mass. The Rashba factor $\alpha$ is chosen as $18.4*10^{-12}$ eV.m, which is a suitable value for these materials [9, 26-27]. Then the reduced Rashba factor is calculated as $\bar{\alpha} = \alpha \frac{m}{\hbar^2} = 0.01$ nm$^{-1}$. The sparse matrix solver UMFPack is used in our calculations.

We set the discretization number $N_1 = 100$ in our calculation. In this paper, we suppose the incident electron wave is polarized in the up direction. The spin polarized electron wave (with the wavelength of 110 nm) goes from the left side into the device. In Fig. 2 (a) we see that in the device region the amplitude of spin up component generally decreases to zero and the amplitude of spin down component generally increases from zero to some fixed value. This is because of the spin precession process [1-2]: in the SOI region the incident electron wave will evolve as the superposition of the SOI eigenmodes. Difference eigenmode has difference group velocity. Thus their interference results in modulated amplitudes for the spin up and down components. From the approximated calculations in the Datta-Das transistor, we have the following expression for the phase difference between the two main eigenmodes in the SOI waveguide [2]: $\Delta\theta = \frac{2m}{\hbar^2}\alpha_z H = 2\bar{\alpha}H$. Substituting $\bar{\alpha}$ =0.01 nm$^{-1}$ and $H$ =155 nm into this expression, we see $\Delta\theta \approx \pi$. This qualitatively explains the transformation of spin up and down components at the exit terminal.

We also calculated the current dependence on the waveguide position in this SOI transistor, as shown in Fig. 2(b). The spin-separated currents are calculated by Eq. (13) and the total current is calculated by the integral of Eq. (25a). Again we see that the spin up current decreases to zero and on the other hand the spin down current increases. But the total current remains a constant. This constant current is reasonable and it can be a benchmark for the HMM calculation.

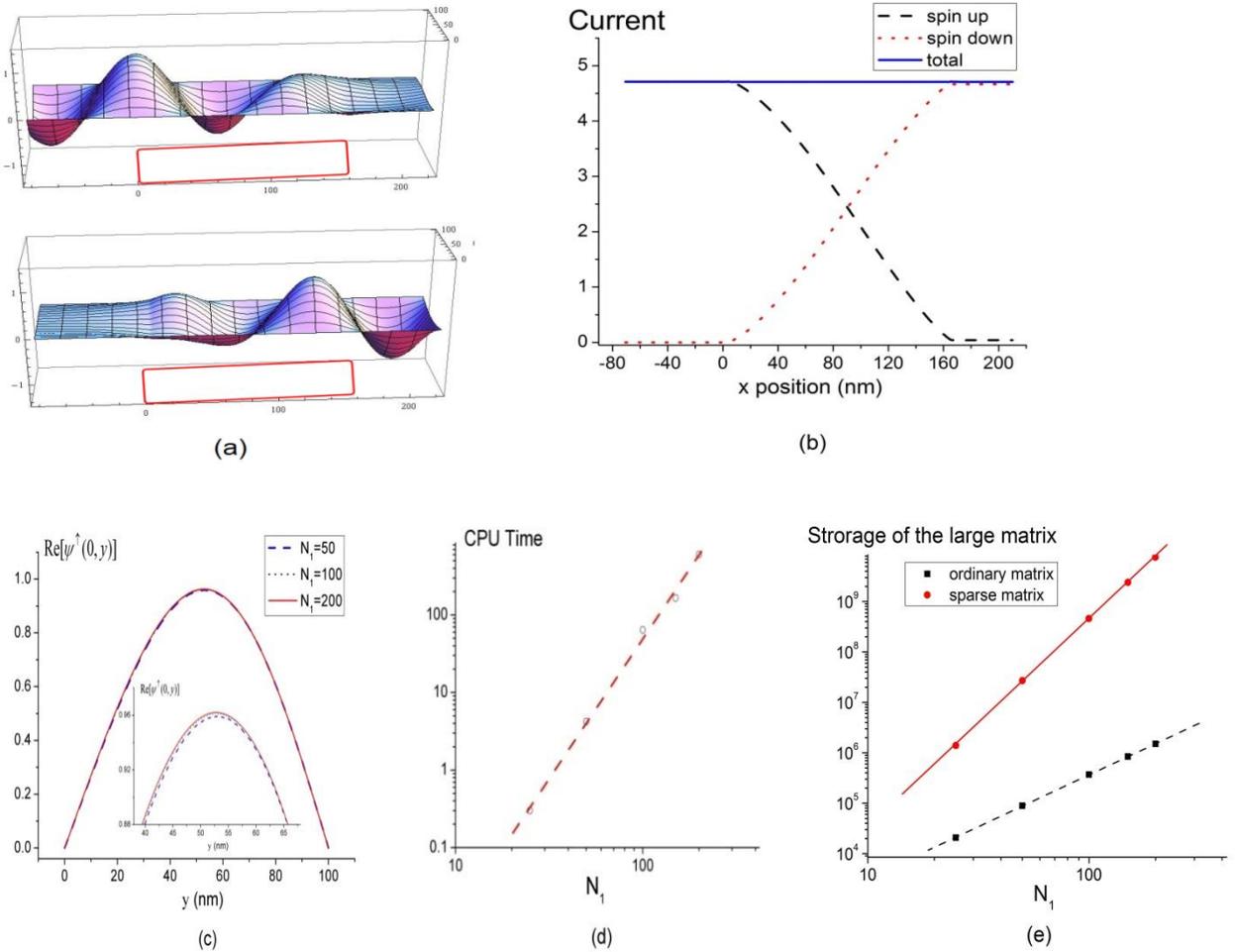

**Fig.2.** (a): The wavefunction (real part) distributions for the spin-up (top) and spin-down (bottom) component of electron in a waveguide with Rashba effect. The red rectangles in the bottom part denote the device regions. (b): The current dependence on the length of the Rashba region given by x. The dashed line and dotted line are for the spin up and down components respectively; the solid line is for the total current calculated from the integral of Eq. (24a). The parameters in this figure are: a=100 nm, $H$=155nm, $\lambda_F$ =10 nm, $\bar{\alpha}$ =0.01 nm$^{-1}$. (c) The real part of wavefunction (spin up component) at the interface of the device and the left lead.

The dashed line, dotted line and the solid line are for $N_1$ =50, 100 and 200 respectively. (d) CPU time dependence on the grid size for the HMM calculations. The dashed line is from the linear fitting. (e) The storage of the large matrix dependence on the grid size in HMM calculation. The two straight lines are from the linear fitting.

## 5.2 Convergence and complexity of the HMM method

Here we investigate the convergence and complexity of our method. The calculation parameters are the same as that in Sec. 5.1, but with different grid size. Figure 2(c) gives the convergence results. The three lines are the wavefunctions (real part and spin up component) at the interface between left lead and device from different discretization. They approach together excellently which shows a good convergence of our HMM calculation.

Figure 2(d) is about the computation time dependence on the grid size. This represents the time complexity of our algorithm. From the linear fitting we see that CPU time is proportional to $N_1^{3.5}$. The main computation time of the HMM method is cost in the sparse matrix solver. It is known that for a sparse direct solver the time complexity is about $O(N^{1.5})$ or $O(N_1^{3.0})$ in a 2D finite-difference grid. (In Sec. 5.1 we see that $N = N_x \propto N_1^2$ for very large $N_1$). Since in our HMM calculations, there some continuity equations (Eqs. (9-12)) which are not as sparse as those in the 2D finite difference case (Eq. 4), so this complexity of $O(N_1^{3.5})$ is reasonable. Fig. 2(d) shows that this sparse matrix technique is much faster than the common matrix solver, which is about $O(N^3)$ or $O(N_1^{6.0})$.

Figure 2(e) is about the space complexity of our method. Due to the usage of the sparse matrix, the memory cost in the calculation is greatly reduced. From the linear fitting in this figure we know that in the sparse matrix case the memory storage for the large matrix is scaled as $O(N_1^2)$ while in the common matrix case it is $O(N_1^4)$. This is because that in the sparse matrix case only the nonzero elements in the large matrix are recorded. So with this technique both the memory and the computation time can be greatly reduced.

## 5.3 Eigenmode calculation in the SOI lead

Now we use the quadratic eigenstate calculations in Section 3.2 for the eigenmodes in SOI waveguides. Figure 3 shows the calculated band structures in different strength of SOI. Figure 3a is for the weak SOI case. There are two branches for each subband due to the Rashba interaction. So for every energy value there are two wavevectors ($k$) in each sub-band. We term them as the 'fast mode' and 'slow mode'. They have different group velocities which results in the spin precession effect. However, for strong SOI case as shown in Fig. 3b, some subband branches get very close and may have anticrossing phenomenon. This was also observed by Moroz and Barnes [7]. In this case we can hardly distinguish the fast or slow modes for each subband.

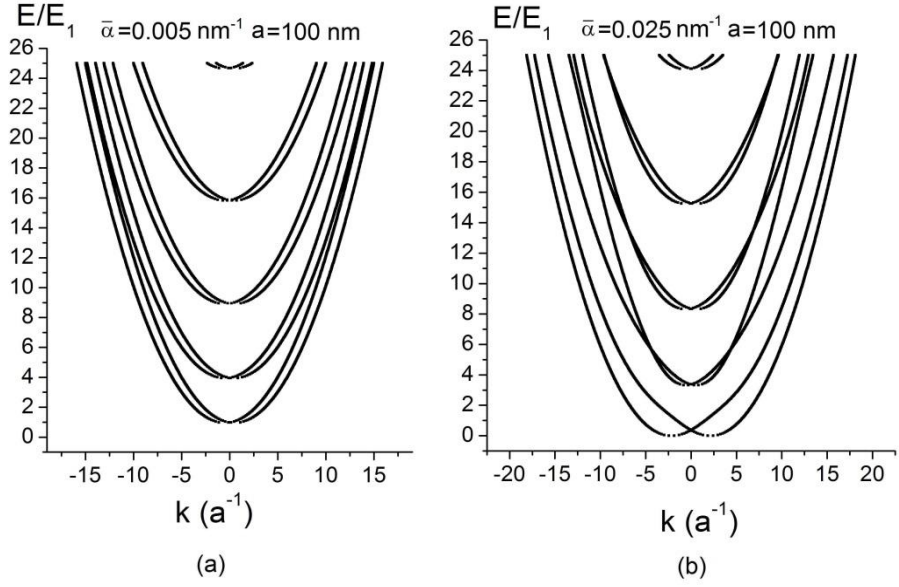

**Fig. 3.** The band structures of uniform Rashba waveguide. (a) is for the weak SOI case ($\bar{\alpha}$ =0.005 nm$^{-1}$) and (b) is for the strong SOI case ($\bar{\alpha}$ =0.025 nm$^{-1}$). The energy is scaled by $E_1$, which is the first subband cut-off energy for the electron waveguide without SOI; the wavevector is scaled by $a^{-1}$, and $a$ =100nm, which is the waveguide width.

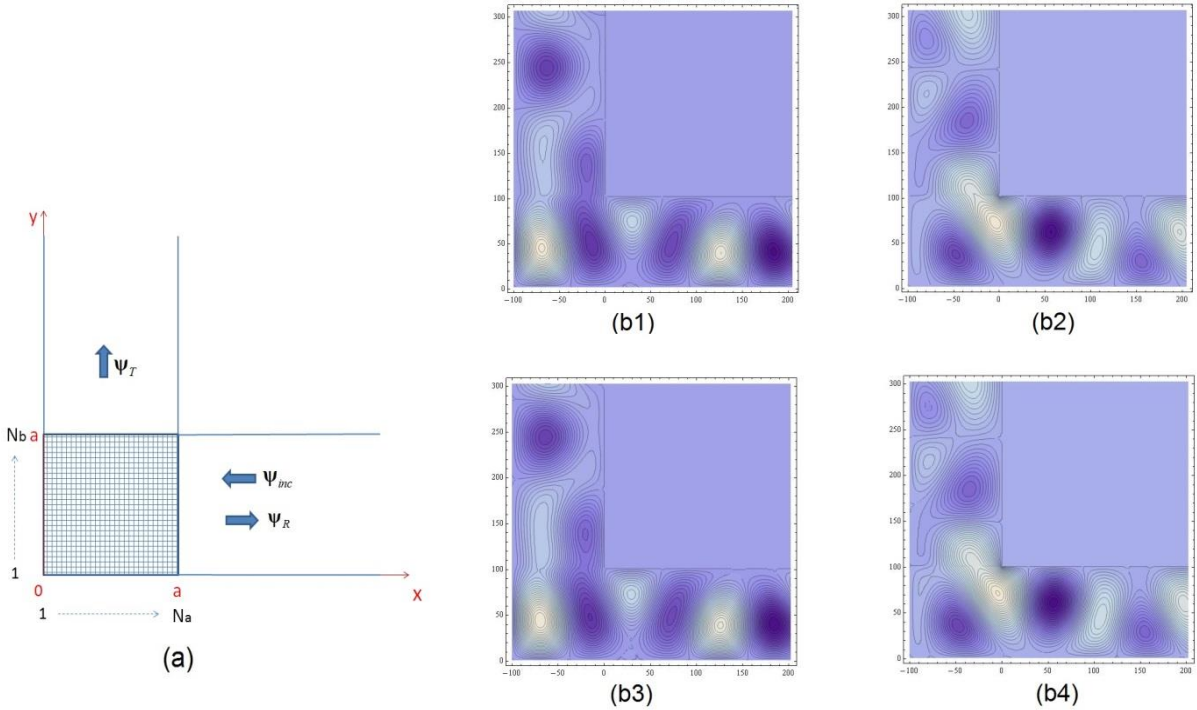

**Fig.4.** (a) The diagram demonstrating the L-shaped bend waveguide for the hybrid boundary-matching method. SOI exists in the device (the meshed corner part) and lead regions. The width of the waveguide is 100 nm; the incident wavelength is 128.9 nm, and the reduce Rashba factor $\bar{\alpha}$ =0.019 nm$^{-1}$. (b1) (spin up component) and (b2) (spin down component) show the wavefunction

distributions (real part) for this system with $N_a=N_b=42$; similarly, (b3) and (b4) show the wavefunction distributions for this system with $N_a=N_b=102$.

## 5.4 HMM calculations for the bend SOI waveguide

After calculating the SOI eigenmodes, we may implement the HMM method for the SOI-lead systems. To emphasis the advantage of this method, we choose an L-shaped bend waveguide for calculation. This type of bending system can hardly be handled by the traditional scattering matrix method [9-10], since it cannot be divided by a series of parallel slices. The bend waveguide is demonstrated in Fig. 4(a). An incident SOI eigenmode ($\psi_{inc}$) is in the lower-right lead. The reflected and transmitted waves ($\psi_R$ and $\psi_T$) in the lower-right and upper-left leads are all expanded into the superposition of such SOI eigenmodes. The meshed part is the device region. We use $N_a=102$ and 42 in our calculations. The continuity equations are set up at the device-lead interfaces. It is noted that the eigenmodes in upper-left lead ($y$ direction) has the form of $\begin{pmatrix} \Phi_m^{\prime\pm,\uparrow}(x) \\ \Phi_m^{\prime\pm,\downarrow}(x) \end{pmatrix} e^{iK_m^{\prime\pm} \cdot y}$.

From Section 3.2, it is easy to get the eigenmodes in $y$ direction from those in $x$ direction. The relations between these two set of SOI eigenmodes are given: $\Phi_m^{\prime\pm,\uparrow} = -i\Phi_m^{\pm,\uparrow}$, $\Phi_m^{\prime\pm,\downarrow} = i\Phi_m^{\pm,\downarrow}$ and $K_m^{\prime\pm} = -K_m^{\pm}$. The hard-wall potential at the device and the leads boundaries are also used in this problem. So due to the special geometry, in the device region, the discretized wavefunction on the nodes of the two leads' sides (at the lines $x=0$ or $y=0$) and that on the 'lead corner' (at the point $x=a$ and $y=a$) are set to be zero. This is different from the basic HMM scheme in Sec. 3.1. The discrete wavefunction has to be indexed differently; also and the number of unknowns is different [28].

Figures 4(b) show the calculated results: the contour of wavefunction (real part). We see the spin-related electron waves are bend according to the geometry of the waveguide. Figure 4(b1) and (b2) are for the mesh size of 42; Figure. 4(b3) and (b4) are for the mesh size of 102. It is easy to see that the two calculations generate very similar results. This ensures the convergence of our SOI-typed HMM calculations.

## 5.5 Transmission calculations for the multiple Fano resonance in the SOI waveguides

Now we use the mode-matching method to investigate the Fano-Rashba resonance and the bound states. We consider a waveguide with non-SOI lead. The device width (a) is 100 nm and the length (H) is 120 nm. As we mentioned before that there often exist a potential well in the Rashba region due to the gate voltage modulation. It is evaluated as $V_0 = -\frac{\hbar^2}{2m}(\frac{2\pi}{\lambda_V})^2$. We choose $\lambda_V = 144.0$ nm, corresponding to $V_0 = -1.7$ meV.

Figure 5(a) shows the transmission spectrum in this 2D waveguide without Rashba interaction. We see for different energy, there are some transmission oscillations. This is due to the Breit-Wigner resonance, resulting from the interference of electron wave between two interfaces of the lowered potential well. From the quantum mechanics textbook, the transmission spectrum in a 1D quantum well is given as [29]

$$T = [1 + \frac{\sin^2(k'H)}{\frac{4E}{V_0}(1+\frac{E}{V_0})}]^{-1}$$

where $k' = \sqrt{\frac{2m^*}{\hbar^2}(E+V_0)}$, $V_0$ is the potential of the quantum well. We see that when $k'H = n\pi$ ($n$ is an integer), the resonance transmission ($T=1$) occurs. In our 2D waveguide, the relation between the transverse and longitudinal wave vectors is: $k'^2 = k_{//}^2 + k_{\perp}^2 = k_{//}^2 + (\frac{\pi}{a})^2$ (The wavefunction is like $e^{\pm ik_1 x}\sin(\frac{\pi y}{a})$ and only the 1$^{st}$ subband is considered). The resonance transmission condition becomes $k_{//}H = n\pi$, where $k_{//} = \sqrt{k'^2 - (\frac{\pi}{a})^2}$. In Fig. 5 (a), we plot the transmission as a function of $k_{//}H$. We see that the transmission peak occurs at $k_{//}H = 2\pi$.

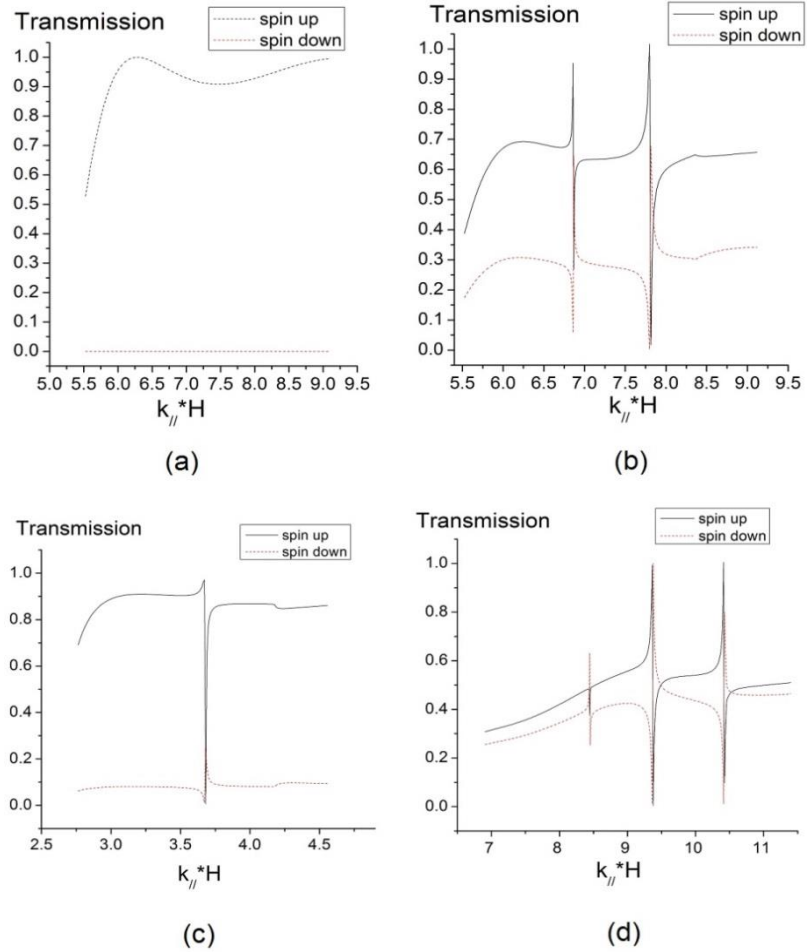

**Fig. 5.** The transmission spectra as a function of $k_{//}H$ for two spin components in some quantum well systems. (a): The system with $H$=120 nm and the device has no Rashba interaction. (b): The system with $H$=120 nm and the device has Rashba interaction. The black solid line is for spin-up component and the red dashed line is for spin-down component. (c): The system with $H$=60 nm and the device has the Rashba interaction. (d): The system with $H$=150 nm and the device has the Rashba interaction. Other parameters are the same: $a$ =100nm; $\bar{\alpha}$ =0.005 nm$^{-1}$, $V_0 = -\frac{\hbar^2}{2m}(\frac{2\pi}{\lambda_V})^2$ with $\lambda_V$ =144 nm.

In the presence of Rashba interaction, both the spin-up and spin-down components appear in the output port due to the spin precession effect, as shown in Fig. 5(b). The very wide peak at around $k_{//}H = 2\pi$ is also observed due to the Breit-Wigner resonance. More importantly, we find that there exist two sets of sharp peaks (and dips) in both transmission curves (The dips lies at $k_{//}H$ = 6.887 and 7.816). This can be explained as the Fano resonance due to the interaction of propagation waves and the bound state modes. The multiple Fano peaks are related to the high-order bound states in the system. We will make a detailed analysis for this phenomenon later.

We also calculate the transmission spectra for the shorter ($H$ = 60 nm) and longer ($H$ = 150 nm) Rashba waveguides, with the same other parameters (such as $\bar{a}$ and $\bar{\alpha}$), as shown in Fig. 5 (c) and Fig. 5(d). We see that there still exist one and three sharp peaks in their spectra respectively. These peaks also reflect the Fano resonance resulting from the interaction of the propagation modes and the bound states.

5.6 Spin bound states in the SOI waveguides

In previous section we hypothesize that the multiple peaks are related to the high-order bound states in the SOI waveguide. To test this assumption, the mode-matching method is employed to calculate these bound states. The detailed description is given in Sec. 4 and here we show the calculation results.

For the SOI waveguide with $H$=120 nm, the energy is chosen between the 1st and 2nd subbands, and in the two lead regions we assume there exist only high-order decaying modes ($\psi = \sin(\frac{n\pi}{a} y)e^{ik_n x}$, $n > 1$). So the wavefunction decays exponentially in the leads to form the bound states. As mentioned in Section 4, the minimum of error ($\Delta$) represents the eigenstate for the bound states. Figure 6(a) shows the curves of $\ln(|\Delta|)$ dependence on $k_{//}H$. The number of eigenmodes (M) in the non-SOI regions is chosen 10 (the dashed line) and 20 (the solid line). These two curves agree very well especially in the dips, which represent good convergence for this eigenstate method. The two dips lie at $k_{//}H$ =6.865 and 7.802. These energies (or $k_{//}H$ values) of the bound states are very close to those of the sharp dips (resonant suppression point) in Fig. 5(b). In the coupling channel theory [16,18], the transmission spectrum has minimum at $E = \varepsilon_0 + \Delta_0 + \delta$, where $\varepsilon_0$ is the energy of the bound state; $\Delta_0$ and $\delta$ are the real parts of $<\phi_0|V_{21}GV_{12}|\phi_0>$ and $\frac{m}{i\hbar^2 kt}<\varphi_l^* V_{12}\phi_0><\phi_0 V_{21}\varphi_r>$. Since in our weak SOI case, $V_{12} \propto \bar{\alpha} \ll 1$, so $\Delta_0 + \delta$ is a very small number, thus the close energies between the Fano dips and the bound states is reasonable.

Figures 6(b1)-6(b4) show the calculated profile of these bound states (the real part of the wavefunction). We see all these states have one node in the transverse direction which is the characteristics of the second subband. In the longitudinal direction, Fig. 6(b1) and Fig. 6(b2) have one extremum which are related to the first bound state; Fig. 6(b3) and Fig. 6(b4) have two extrema which are related to the second bound state.

Our mode-matching calculations also show that for the short ($H$=60 nm) and long ($H$=150 nm) waveguides, there still exist such bound states. For the short one there exist the 1st order bound state and for the long one there exist bound state of the orders 1- 3. All these bound states have the similar energies with the Fano peaks as in Fig. 5 (c) and Fig. 5(d). So we can confirm that these multiple sharp peaks come from the high-order Fano resonance.

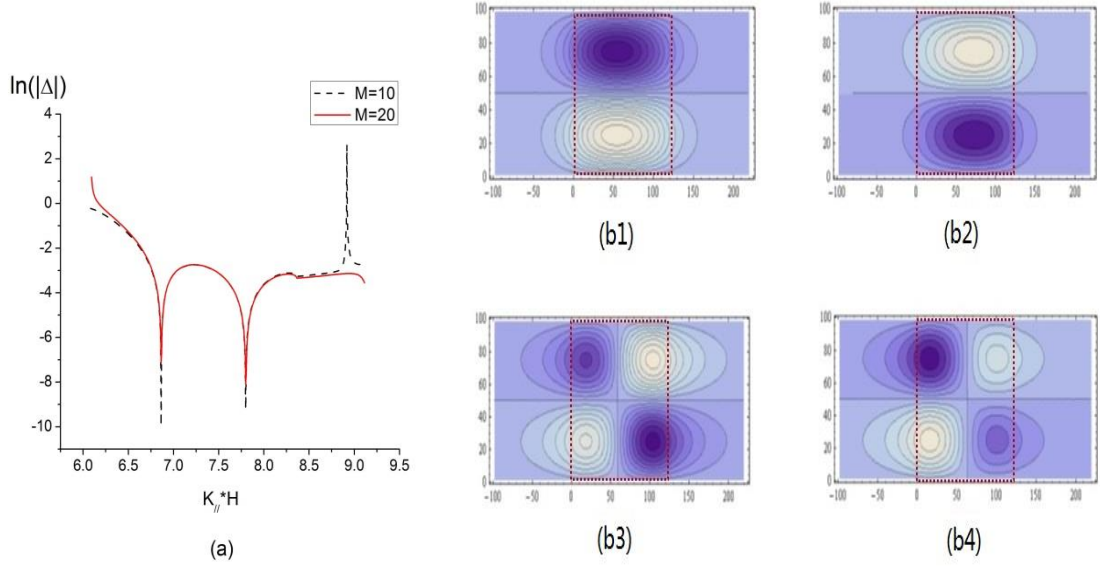

**Fig. 6.** (a) The $\ln(|\Delta|)$ v.s. $k_{//}H$ relation in the eigenvalue calculation for the bound states of a Rashba quantum well system (*H*=120 nm). The two dips denote the eigenvalue energies for the bound states. The dashed line is for M=10 and the solid line is for M=20. (b1)-(b4) are the calculated eigenmodes of the spin bound states. (b1) and (b2) are for the first-order bound state; (b3) and (b4) are for the second-order bound state. The left and right columns are for the spin-up and spin-down components respectively. All the other parameters are as the same as those in Fig. 5.

## 6. Conclusion

We have developed the hybrid mode-matching method for the spin-dependent transport calculations. The system is about the 2D electron gas in the Rashba waveguide on the semiconductor hetero-structure. Both the non-SOI lead and SOI lead have been considered. For the latter case the SOI eigenmodes have been solved from the mode expansion and quadratic eigenproblem schemes. We also have investigated the spin bound states in the Fano-Rashba resonance with a simplified mode-matching method. The calculated eigen-energies of these bound states are consistent with the transmission spectra of the Rashba quantum-well systems.

**Acknowledgements**

The authors greatly thank for the useful and enlightening discussions with Prof. Ping Sheng in the department of physics, the Hong Kong University of Science and Technology. The useful suggestions and helps from Dr. Yan Zhou in the department of physics from the University of Hong Kong are also acknowledged.